\documentstyle[preprint,aps]{revtex}
\begin{document}
\preprint{UH-511-802-94, OITS-550}
\draft
\title{ CP VIOLATION IN HYPERON DECAYS\footnote{Talk presented by Xiao-Gang He
at the Eighth Meeting of the American Physical Society, Division of Particles
and Fields (DPF'94), Albuqurque, New Mexico, August 2-6, 1994}}
\author{Xiao-Gang He$^a$ and Sandip Pakvasa$^b$}
\address {$^a$Institute of Theoretical Science\\
University of Oregon\\
Eugene, OR 97403-5203, USA\\
and\\
$^b$Department of Physics and Astronomy\\
University of Hawaii\\
Honolulu, HI 96822, USA}
\date{August, 1994}
\maketitle
\begin{abstract}
The present status for CP violation in hyperon decays is reviewed.
\end{abstract}
\newpage

\section{Introduction}

Non-leptonic hyperon decays of $\Lambda$, $\Sigma$ and
$\Xi$\cite{1,2,3,4,5,6,7,8} are interesting processes to test CP conservation
outside the neutral Kaon sysytem. Measurements of CP violation in hyperon
decays will provide us with useful information about the origin of CP
violation. Several proposals have been made to
look for CP violation in hyperon decays\cite{6,7}. Recently the E871 proposal
at Fermilab has been approved\cite{8}. The
expected sensitivity for CP violation test is about the same order of magnitude
as the Standard Model (SM) prediction. This experiment will measure CP
violation in
$\Xi^- \rightarrow \Lambda \pi^-$ and $\Lambda \rightarrow p\pi^-$. The E871
experiment will start to take data as early as 1996.
 In this talk we will review the present status for CP violation in hyperon
decays. We will concentrate on $\Xi$ and $\Lambda$ decays because there is not
hope to measure
CP violation in $\Sigma$ decays in the near future.

Non-leptonic hyperon decays can proceed into both S-wave and P-wave final
states with amplitudes S and P, respectively. One can write the amplitude as
\begin{eqnarray}
Amp(B_i \rightarrow B_f\pi) =S + P\vec \sigma\cdot \vec q\;,
\end{eqnarray}
where $\vec q$ is the momentum of the final baryon $B_f$.
Experimental observables are: the decay width $\Gamma$, and the parameters in
the decay angular distribution. In the rest frame of the initial hyperon, the
angular distribution is given by
\begin{eqnarray}
{4\pi\over \Gamma}{d\Gamma \over d\Omega} = 1 +\alpha \hat s_i\cdot \hat q +
\hat s_f \cdot [(\alpha + \hat s_i\cdot \hat q)\hat q +\beta \hat s_i\times
\hat q + \gamma(\hat q\times (\hat s_i\times \hat q))]\;,
\end{eqnarray}
where $s_{i,f}$ are the spins of the initial and final baryons, respectively.
 $\hat v $ indicates the direction of the corresponding vector. The parameters
$\alpha\;, \beta$ and $\gamma$ are defined as
\begin{eqnarray}
\alpha = {2Re(S^*P)\over |S|^2+|P|^2}\;, \;\;\beta = {2Im(S^*P)\over
|S|^2+|P|^2}\;,\;\;\gamma &=&{|S|^2-|P|^2\over |S|^2+|P|^2}\;.
\end{eqnarray}
Only two of them are independent. We will discuss $\alpha$ and $\beta$. In the
literature, $\beta$ is sometimes
parametrized as $\beta = \sqrt{1-\alpha^2} sin\phi$.

It is convenient to write the amplitudes
as
\begin{eqnarray}
S = \sum_i S_i e^{i(\phi^S_i + \delta_i^S)}\;,\;\;P = \sum_i P_i e^{i(\phi^P_i
+ \delta_i^P)}\;
\end{eqnarray}
to explicitly separate the strong rescattering phases $\delta_i$ and the weak
CP
violating phases $\phi_i$.

The decay amplitudes $\bar S$ and $\bar P$ for anti-hyperon can be parametrized
in a similar way. Then
\begin{eqnarray}
\bar S = -\sum_i S_i e^{i(-\phi^S_i + \delta_i^S)}\;,\;\;\bar P = \sum_i P_i
e^{i(-\phi^P_i + \delta_i^P)}\;.
\end{eqnarray}
We will denote the observables in anti-hyperon decays with a bar on the
corresponding ones in hyperon decays.

\section{ CP Violating Observables}

Several CP violating observables can be constructed using the observables
discussed in the previous section. The interesting ones are\cite{2}
\begin{eqnarray}
\Delta &=& {\Gamma - \bar \Gamma \over \Gamma +\bar \Gamma}\;,\;\;
A = {\Gamma \alpha +\bar \Gamma \bar \alpha \over \Gamma \alpha -\bar \Gamma
\bar \alpha } \approx {\alpha +\bar \alpha\over \alpha-\bar \alpha} + \Delta\;,
\nonumber\\
B &=& {\Gamma \beta +\bar \Gamma \bar \beta \over \Gamma \beta -\bar \Gamma
\bar \beta } \approx {\beta +\bar \beta\over \beta-\bar \beta} + \Delta\;.
\end{eqnarray}

All these CP violating observables can, in principle, be measured
experimentally.
It has been shown that the low energy reaction $p_i\bar p_i \rightarrow \Lambda
\bar \Lambda \rightarrow p_f\pi^- \bar p_f \pi^+$ can be used to measure A for
$\Lambda$\cite{9}. The measurement
\begin{eqnarray}
\tilde A = {N_p^+ -N_p^- +N^+_{\bar p} - N^-_{\bar p} \over N_{total}}\;,
\end{eqnarray}
is equal to $\it P_\Lambda \alpha_\Lambda A(\Lambda)$. Here
$N^{\pm}_p$ indicates events with $(\hat p_i\times \hat P_\Lambda) \cdot \hat
p_f  > 0$ or $<0$, and similarly for anti-particles. $\it P_\Lambda$ is the
polarization of the $\Lambda$ produced in the $p\bar p$ collision.

The measurement of $B$ requires the analysis of the polarization of the
final baryon. Low energy $p \bar p$ collision can also measure $B$ for $\Xi$
decays\cite{9}. In the process, $p_i\bar p_i \rightarrow \Xi \bar \Xi
\rightarrow \Lambda \pi^- \bar \Lambda \pi^+ \rightarrow
p_f\pi^- \pi^- \bar p_f \pi^+ \pi^+$, one can measure
\begin{eqnarray}
\tilde B = {\tilde N_p^+ - \tilde N_p^- + \tilde N^+_{\bar p} - \tilde
N^-_{\bar p} \over N_{total}}\;,
\end{eqnarray}
where $\tilde N^{\pm}_p$ indicates events with $\hat P_\Xi \cdot (\hat p_f
\times \hat
p_\Lambda) > 0$ or $<0$, and similarly for anti-particles. $\tilde B$
is given by $(\pi/8) \it P_\Xi \alpha_\Lambda \beta_\Xi ( A(\Lambda) +
B(\Xi))$,
and hence a measurement of $\tilde A$ and $\tilde B$ yields $A(\Lambda)$ and
$B(\Xi)$.

There have been some measurements for CP violation in hyperon decays by several
groups,
\begin{eqnarray}
A(\Lambda \rightarrow p\pi^-)
&=& -0.02 \pm 0.14\;,\;\; \mbox{From} \; p\bar p \rightarrow \Lambda X\;
\mbox{and} \; p \bar p
\rightarrow \bar \Lambda X. \;\;\;\;\mbox{Ref.[10]}\nonumber\\
A(\Lambda \rightarrow p\pi^-)
&=& -0.07 \pm 0.09\;,\;\; \mbox{From}\;p\bar p \rightarrow \Lambda \bar\Lambda
\rightarrow p\pi^-\bar p \pi^+. \;\;\;\;\;\;\;\;\;\mbox{Ref.[11]}\nonumber\\
A(\Lambda \rightarrow p\pi^-)
&=& 0.01 \pm 0.10\;, \;\;\;\;\;\mbox{From}
\;J/\psi \rightarrow \Lambda \bar\Lambda
\rightarrow p\pi^-\bar p \pi^+.\;\;\;\;\;\;\mbox{Ref.[12]}\nonumber\\
\end{eqnarray}

The experiment at E871 will measure $\alpha_\Lambda \alpha_\Xi$ in the
decay $\Xi^- \rightarrow \Lambda \pi^-\rightarrow p\pi^-\pi^-$, and similar
measurement for anti-$\Xi$ decays\cite{8}. An asymmetry $A_{asy}$ can be
extracted
\begin{eqnarray}
A_{asy} = {\alpha_\Lambda\alpha_\Xi - \bar \alpha_\Lambda \bar \alpha_\Xi
\over \alpha_\Lambda\alpha_\Xi + \bar \alpha_\Lambda \bar \alpha_\Xi}\approx
A(\Lambda) + A(\Xi)\;.
\end{eqnarray}
The expected sensitivity for $A_{asy}$ is $10^{-4}$ and may reach $10^{-5}$
which would test the SM predictions.

\section{Theoretical Calculations}

There are large uncertainties in theoretical calculations for the CP violating
observables due to our poor understanding of the hadronic matrix elements. To
reduce errors it is best to use experimental measurements for CP conserving
quantities and to calculate CP violating parameters theoretically, that is, we
calculate the weak phases $\phi^{s,p}$. The experimental data on CP conserving
quantities are summarized below.

The isospin decomposition of $\Lambda$ and $\Xi$ are given by\cite{13}
\begin{eqnarray}
\Lambda \rightarrow p\pi^-\;,\;\;
S(\Lambda^0_-) &=& -\sqrt{{2\over 3}} S_{11}e^{i(\delta_1 +\phi^s_1)}
+\sqrt{{1\over 3}}S_{33} e^{i(\delta_3+ \phi^s_3)})\;,\nonumber\\
P(\Lambda^0_-) &=& -\sqrt{{2\over 3}} P_{11}e^{i(\delta_{11} +\phi^p_1)}
+\sqrt{{1\over 3}}P_{33} e^{i(\delta_{33}+ \phi^p_3)})\;,\nonumber\\
\Lambda \rightarrow n\pi^0\;,\;\;
S(\Lambda^0_0) &=& \sqrt{{1\over 3}} S_{11}e^{i(\delta_1 +\phi^s_1)}
+\sqrt{{2\over 3}}S_{33} e^{i(\delta_3 +\phi^s_3)})\;,\nonumber\\
P(\Lambda^0_0) &=& \sqrt{{1\over 3}} P_{11}e^{(i\delta_{11} +\phi^p_1)}
+\sqrt{{2\over 3}}P_{33} e^{i(\delta_{33}+ \phi^p_3)})\;,\nonumber\\
\Xi^- \rightarrow \Lambda\pi^-\;,\;\;
S(\Xi_-^-) &=& S_{12}e^{i(\delta_2 +\phi^s_{12})}
+{1\over 2}S_{32} e^{i(\delta_2 +\phi^s_{32})})\;,\nonumber\\
P(\Xi_-^-) &=& P_{12}e^{i(\delta_{21} +\phi^p_{12})}
+{1\over 2}P_{32} e^{i(\delta_{21}+\phi^p_{32})})\;,\nonumber\\
\Xi^0 \rightarrow \Lambda\pi^0\;,\;\;
S(\Xi_-^-) &=& \sqrt{{1\over 2}}(S_{12}e^{i(\delta_2 +\phi^s_{12})}
-S_{32} e^{i(\delta_2+\phi^s_{32})})\;,\nonumber\\
P(\Xi_-^-) &=& \sqrt{{1\over2}}(P_{12}e^{i(\delta_{21} +\phi^p_{12})}
-P_{32} e^{i(\delta_{21}+\phi^p_{32})})\;.
\end{eqnarray}

These decays are dominated by the $\Delta I = 1/2$ amplitudes. Experimental
measurements give:\cite{14}
$S_{33}/S_{11} = 0.027\pm 0.008$, $P_{33}/P_{11} = 0.03 \pm 0.037$;
$S_{32}/S_{12} = -0.046 \pm 0.014$, and $P_{32}/P_{12} = -0.01 \pm 0.04$. From
$N\pi$ scattering, the
strong rescattering phase for $\Lambda$ decays are determined to be\cite{15}: $
\delta_1 \approx 6.0^0$,
$\delta_3 \approx -3.8$, $\delta_{11} \approx -1.1^0$ and $\delta_{31} =
-0.7^0$
with errors of order $1^0$. The strong rescattering phases for $\Xi$ decays are
not exprimentally determined. Theoretical
predictions very a large range. Nath and Kumer\cite{16} obtained:
$\delta_{21} = -2.7^0$ and $\delta_2=-18.7$. Martin\cite{17} obtained
$\delta_{21} = -1.2^0$.
Recently, Lu, Savage and Wise\cite{18},using chiral pertubation theory,
obtained
$\delta_{21} = -1.7^0$ and $\delta_2 = 0$ to the lowest order. In this last
estimate, contributions from $1/2^-$ and $3/2^-$ states are not included, which
can give rise to a significant $\delta_2$.

To a very good approximation, the CP violating observables can be simplified
to yield\cite{2},
\begin{eqnarray}
\Delta (\Lambda^0_-) &=& -2\Delta (\Lambda^0_0) = \sqrt{2}
{S_{33}\over S_{11}} \mbox{sin}(\delta_3 - \delta_1)\mbox{sin}(\phi^s_3
-\phi^s_1)\;,\nonumber\\
A(\Lambda^0_-) &=& A(\Lambda^0_0) =
-\mbox{tan}(\delta_{11}-\delta_1)\mbox{sin}(\phi^p_1-\phi^s_1)\;,\nonumber\\
B(\Lambda^0_-) &=& B(\Lambda^0_0) =
\mbox{cot}(\delta_{11}-\delta_1)\mbox{sin}(\phi^p_1 -\phi^s_1)\;,\nonumber\\
\Delta (\Xi^-_-) &=& \Delta(\Xi^0_0) = 0\;,\nonumber\\
A(\Xi^-_-) &=& A(\Xi^0_0) =
-\mbox{tan}(\delta_{21}-\delta_2)\mbox{sin}(\phi^p_{12}-\phi^s_{12})\,\nonumber\\
B(\Xi^-_-) &=& B(\Xi^0_0) =
\mbox{cot}(\delta_{21}-\delta_2)\mbox{sin}(\phi^p_{12} -\phi^s_{12})\;.
\end{eqnarray}
 It is well known that for large top quark mass,
there is considerable cancellation for the $I=0$ and $I=1$ contributions
to $\epsilon'/\epsilon$, and $\epsilon'/\epsilon$ can be quite small\cite{19}.
Such cancellation does not happen to the quantities $A$ and $B$ because they
are dominated by $I=1/2$ quantities. Hence hyperon decays probe a somewhat
different operators, even in the SM, from $\epsilon'/\epsilon$.

\subsection{ The Standard Model Predictions.}

In the SM, the origin of CP violation is the non-trivial phase in the
KM matrix\cite{20}. The effective Hamiltonian responsible for non-leptonic
hyperon decays
is given by
\begin{eqnarray}
H_{eff} = {G_F\over \sqrt{2}} V_{ud}^*V_{us} \sum_i C_i(\mu) Q_i\;,
\end{eqnarray}
where the sum is over all the $Q_i$ four-quark operators, and the $C_i(\mu)$ is
the Wilson coefficients. $C_i$ contains both the CP conserving and CP violating
parts. To separate KM mixings and other dependences, $C_i$ is usually
parametrized
as $C_i = z_i + \tau y_i$, where $\tau = -V_{td}^*V_{ts}/V_{ud}^*V_{us}$. CP
violating part is proportional to $Im(\tau)$. To obtain the weak phases, we
need to evaluate
\begin{eqnarray}
Im\it M = {G_F\over \sqrt{2}} V_{ud}^*V_{us}Im(\tau)<\pi B_f|\sum_i y_i(\mu)
Q_i(\mu)|B_i>\;.
\end{eqnarray}

The quantity $y_i(\mu)$ is calculated by taking $y(m_W)$ as an initial value
and then using renormalization group equation to reach the scale
$\mu$\cite{19}. The most difficult part of the calculation is to evaluate $<\pi
B_f|Q_i|B_i>$. At present, there is no convincing method to calculate these
matrix elements.
There are many models which can give estimates. However, it is known that all
these
models can not satisfactorily explain the $I=1/2$ dominance in hyperon decays.
They can at most
produce the experimental amplitudes up to a factor of 2. It is therefore
expected that the estimate for CP violation in hyperon decays can easily off
by a factor of 2. The following is the result obtained by using the
vacuum saturation and factorization approximation\cite{4},
\begin{eqnarray}
Im \it M = {G_F\over \sqrt{2}} V_{ud}^*V_{us} Im(\tau)[(M^s_1 +M_3^s)\it V
+(M_1^p+M_3^p)\it P]\;,
\end{eqnarray}
with
\begin{eqnarray}
M^s_1 &=& {y_1-2y_2\over 3} - {y_7\over 2} +\xi ({-2y_1+y_2\over 3}
-y_3 - {y_8\over 2} ) - Y (y_6 + {y_8\over 2}
+\xi(y_5 +{y_7\over 2}))\;,\nonumber\\
M^p_1 &=& {y_1-2y_2\over 3} - {y_7\over 2} +\xi ({-2y_1+y_2\over 3}
-y_3 - {y_8\over 2}) + Z(y_6 + {y_8\over 2}
+\xi(y_5 +{y_7\over 2}))\;,\nonumber\\
M^s_3 &=& - {y_1+y_2\over 3}(1+\xi) +{y_7\over 2}+{y_8\over 2}- Y (\xi y_7 +
y_8)\;,\nonumber\\
M^p_3 &=& - {y_1+y_2\over 3}(1+\xi) +{y_7\over 2} + {y_8\over 2}+ Z (\xi y_7 +
y_8)\;,
\end{eqnarray}
where $Y = 2m_\pi^2/(m_u+m_d)(m_s-m_u)$, $Z =2m_\pi^2/(m_u+m_d)(m_s+m_u)$, and
$\xi = 1/N$ with N the number of color.

For $\Lambda \rightarrow p\pi^-$,
\begin{eqnarray}
V = i \sqrt{2}F_\pi (m_\Lambda - m_p)\sqrt{{3\over 2}}\bar p \Lambda\;,\;\;P
\approx -i {2F_\pi F_K m_\pi^2\over m_K^2-m_\pi^2} g_{\Lambda pK} \bar p
\gamma_5 \Lambda\;,
\end{eqnarray}
where $m_i$ are the masses of the particle i, $F_\pi = 93$ MeV, $F_K \approx
1.3 F_\pi$ and $g_{\Lambda pK} \approx -13.3$.
Similarly one can obtain the decay apmlitudes for $\Xi$ decays.

Using these estimates for the matrix elements and information for the CP
violating parameter $Im(\tau)$ from $\epsilon$ and other constraints\cite{4},
predictions for the CP violating observables $\Delta$, A and B can be obtained.
$\Delta$ is predicted to be  less than $10^{-6}$. It is very small. The
parameter
$A(\Lambda)$ is in the range $-(0.5 \sim 0.1)\times 10^{-4}$. $B(\Lambda)$ is
about 60 times larger.

The same calculation has been done using MIT bag model\cite{2} and other models
for hadronic matrix elements\cite{3,4}.
The MIT bag model predicts the same orders of magnitude for $A$ and $B$ as the
vacumm saturation predictions. Larger values are possible in other
models\cite{4}. It has recently been shown that the gluon dipole operator also
has significant contributions to
$A(\Xi)$\cite{5}. In Table 1, We list the
allowed ranges of $\Delta $, $A$ and $B$ using MIT bag model for $\Lambda$ and
$\Xi$ decays. The ranges include uncertainties from the KM matrix elements and
uncertainties in top quark mass\cite{4}. In particular, $A(\Xi)$ is in the
range $-(0.1 \sim 1)\times 10^{-4}$ and hence the quantity $A_{asy}$ to be
measured by
E871, $A(\Lambda) + A(\Xi)$, is expected to be in the range $-(0.2\sim
1.5)\times 10^{-4}$.

\subsection{The Multi-Higgs Model Predicitions.}

I will consider multi-Higgs model with neutral flavour current conservation at
the tree level
and CP is violated spontaneously. This is the model proposed by
Weinberg\cite{21}. In this model the most important operator related to CP
violation is
\begin{eqnarray}
L_{CPV} = i \tilde f \bar d T^a\sigma_{\mu\nu} (1-\gamma_5) s G^{\mu\nu}_a\;,
\end{eqnarray}
where $G^{\mu\nu}_a$ is the gluon field strength, $T^a$ is the $SU(3)_C$
generator, and $\tilde f$ is a constant depending
on several parameters. This operator can reproduce CP violation in the neutral
Kaon sector provided that\cite{22}
\begin{eqnarray}
m_K\sqrt{2} |\epsilon| \Delta m_{K_L-K_S} \approx 10^{-7} <
\pi^0|L_{CPV}|K^0>\;.
\end{eqnarray}
This fixes the strength of CP violation in this model. The predictions for
CP violation in hyperon decays have been carried out in Ref.[2] using bag model
and pole model calculations. The results are listed in Table 1.

In models in which flavor changing neutral currents are responsible for CP
non-conservation, all effect in hyperon decays as well as $\epsilon'/\epsilon$
are essentially zero.

\subsection{ The Left-Right Symmetric Model Predictions.}

The Left-Right symmetric models are based on the gauge group $SU(3)_C\times
SU(2)_L
\times SU(2)_R\times U(1)_{B-L}$. In this model there are additional CP
violating phases from the right-handed KM matrix. Here we consider a simple
model of
this type, the "isoconjugate" Left-Right model\cite{23}. In this model there is
no mixing between $W_L$ and $W_R$. There is
no CP violation in the left-handed sector. All CP violations are coming from
the
right-handed KM matrix. And $|V_{Lij}| = |V_{Rij}$ for the KM matrices. The
full $\Delta S = 1$ Hamiltonian has the form
\begin{eqnarray}
H_W = {G_F\over \sqrt{2}} (O_{LL} + \eta e^{i\beta} O_{RR})\;.
\end{eqnarray}
The operators $O_{LL}$ and $O_{RR}$ are identical operators, except that
$O_{LL}$ is a product of two left-handed currents whrease $O_{RR}$ has two
right-handed currents. Because this structure, one can easily see that
parity-nonconserving processes have an identical phase factor $1+i\eta\beta$,
while all parity-conserving ones have phase $1-i\eta\beta$. We have:
$\phi^s_i = \eta\beta$ and $\phi^p_i = -\eta\beta$ for all decays. The strength
is fixed by requiring this phase to explain CP violation in the neutral Kaon
system\cite{24}. From this consideration, $\eta\beta$ is determined to be about
$4.4\times 10^{-5}$. Because the simple phase structure, $\Delta$ is always
zero in this model. The predictions for $A$ and $B$ are given in Table 1.

\vspace{5mm}
Table 1. Predictions of CP violation in hyperon decays.

\begin{tabular}{|cccc|}
\hline
$\Lambda\;\mbox{decay}$&KM model&Weinberg Model& Left-Right Model\\ \hline
$\Delta(\Lambda^0_-)$& $<10^{-6}$&$-0.8\times 10^{-5}$&0\\ \hline
$A(\Lambda^0_-)$& $-(5\sim 1)\times 10^{-5}$& $-2.5\times 10^{-5}$&$-1.1\times
10^{-5}$\\ \hline
$B(\Lambda^0_-)$& $(3\sim 0.6)\times 10^{-4}$&
$1.6\times 10^{-3}$&$7.0\times 10^{-4}$\\ \hline
$\Xi\; \mbox{decay}$&&&\\ \hline
$\Delta(\Xi^-_-)$& 0&0&0\\ \hline
$A(\Xi^-_-)$&$-(10\sim 1)\times 10^{-5}$&$-3.2\times 10^{-4}$&$2.5\times
10^{-5}$\\ \hline
$B(\Xi^-_-)$&$(10\sim 1)\times 10^{-3}$&$3.8\times 10^{-3}$&$-3.1\times
10^{-4}$\\ \hline
\end{tabular}
\vspace{5mm}

{}From Table 1 we see that,in general, $\Delta$ is very small. It may be
difficult to measure it experimentally. The prediction for the CP violating
observable $A$ is
close to the region which will be probed by the E871 experiment, with
experimental sensitivity $10^{-4}$ to $10^{-5}$ in $A_{asy}$,
new and useful information about CP violation will be obtained.

We thank Drs. N. Deshpande J. Donoghue, H. Steger and G. Valencia for
collaboratotions on the related topic. This work was supported in part by the
U.S. D.O.E. under grant DE-FG06-85ER40224 and DE-FG03-94ER40833.

\end{document}